\documentclass[showpacs,twocolumn,preprintnumbers,showkeys,superscriptaddress,amsmath,amssymb,nofootinbib]{revtex4-2}
\usepackage{color}
\usepackage{amsmath}
\usepackage{amssymb}
\usepackage[utf8]{inputenc}
\usepackage{amsfonts}
\usepackage{bm}
\usepackage{bbold}
\usepackage{graphics}
\usepackage{graphicx}

\newcommand{\be}{\begin{equation}}
\newcommand{\ee}{\end{equation}}
\newcommand{\ba}{\begin{eqnarray}}
\newcommand{\ea}{\end{eqnarray}}

\def\ni{\noindent}

\begin{document}
%
%

\title{\Large Considerations on anomalous photon and $Z$-boson
     self-couplings from the Born-Infeld weak hypercharge action}



\author{M. J. Neves} \email{mariojr@ufrrj.br}
\affiliation{Departamento de F\'{i}sica, Universidade Federal Rural do Rio de Janeiro, BR 465-07, 23890-971, Serop\'edica, RJ, Brazil}

\author{ L. P. R. Ospedal } \email{leoopr@cbpf.br}
\affiliation{Centro Brasileiro de Pesquisas F\'isicas, Rua Dr. Xavier Sigaud
150, Urca, Rio de Janeiro, Brazil, CEP 22290-180}

\author{J. A. Helay\"el-Neto}\email{helayel@cbpf.br}
\affiliation{Centro Brasileiro de Pesquisas F\'isicas, Rua Dr. Xavier Sigaud
150, Urca, Rio de Janeiro, Brazil, CEP 22290-180}

\author{Patricio Gaete} \email{patricio.gaete@usm.cl}
\affiliation{Departamento de F\'{i}sica and Centro Cient\'{i}fico-Tecnol\'ogico de Valpara\'{i}so-CCTVal,
Universidad T\'{e}cnica Federico Santa Mar\'{i}a, Valpara\'{i}so, Chile}

\date{\today}

\begin{abstract}
\ni

We investigate the effects of the Born-Infeld action on the Abelian sector of the electroweak model. The consequence of this approach is the emergence of anomalous couplings in the neutral sector of the $Z$-gauge boson and photon. These new couplings consist of quartic interactions of the photon with the $Z$-particle, as for example, three-photon-and-one-$Z$ vertex. With that, we obtain the decay width of $Z \to 3\,\gamma$ from which we impose a bound on the Born-Infeld parameter. Other bounds are also obtained from the photon quartic couplings.
Subsequently, we consider the presence of an external uniform magnetic field in connection with this Born-Infeld weak hypercharge model. The magnetic background field yields new kinematic effects, like the kinetic mixing between the photon and the $Z$-boson, and we obtain thereby the corresponding dispersion relations for the mixed photon-$Z$-particle system. Finally, we calculate the lowest-order modifications to the interaction energy for the anomalous coupling $3\gamma-Z$, within the framework of the gauge-invariant but path-dependent variables formalism. Our results show that the interaction energy contains a linear term leading to the confinement of static probe charges. With the help of the potential that comes out, interparticle forces are estimated.

\end{abstract}




\maketitle

\pagestyle{myheadings}
\markright{Considerations on anomalous photon and $Z$-boson
     self-couplings from the Born-Infeld weak hypercharge action}


\section{Introduction}
%

%
%

%

One of the most actively pursued areas of research in Particle Physics
consists of the investigation of extensions of the Standard Model (SM) of fundamental interactions. However, the need for consistent extensions have been suggested to overcome a number of theoretical difficulties,
such as the hierarchy problem, the origin of dark matter and the dynamic origin of the Higgs mechanism \cite{BriefingBook,Kasakov2019}.

In this connection, it may be recalled that the electroweak sector of the SM, based on the $SU_L(2) \times U_Y(1)$ Yang-Mills gauge group, yields the three- and four-gauge boson couplings, $W^+ \, W^- \, Z$, $W^+ \, W^- \, \gamma$, $W^+ \, W^- \, \gamma \, \gamma$, $W^+ \, W^- \, Z \, Z$, $W^+ \, W^- \, \gamma \, Z$ and $W^+ \, W^- \, W^+ \, W^-$, as vertices of the classical Lagrangian density. Though the ATLAS and CMS Collaborations of the Large Hadron
Collider (LHC) show a good agreement with the Standard Model predictions, and confirm the structure of three- and four-vertices,
the electroweak symmetry rules out direct couplings with three and four neutral gauge bosons at tree-level. The scrutiny of the electroweak
gauge boson self-interactions is an important task in connection with the breaking mechanism of the electroweak symmetry and, in this
direction, it is known that LHC data set strict limits on the triple and quartic couplings amongst the gauge bosons \cite{Kubota, Kupco,ATLAS2017,Aaboud,CMS2018}.


In this way, any departure from the SM behavior would therefore give a positive signature for the existence of a new and unconventional
physics, including rare SM decay processes \cite{AtlasCollab,Novikov,Dong,Yang, Aaltonen,Baillargeon,Pham,Inan}. Mention should be made, at this point, to the rare decay $Z \to 3 \,\gamma$, which can  be obtained within the SM via one-loop (or higher) Feynman diagrams. Let us also recall here that the investigation of quartic neutral couplings
is justified because these anomalous couplings would point to some new physics beyond the SM. In fact, our present-day expectation is that the LHC TeV scale may be sensitive to effects of this new physics, and, in the ATLAS and CMS Collaborations, the search for anomalous gauge boson couplings is carried out in a model-independent way.

It is worthy recalling, at this stage, that an Effective Field Theory (EFT) framework is a good way to study rare couplings. The advantage of using this development lies in the fact that the gauge symmetries of the SM are respected. To be more precise, electroweak anomalous gauge couplings may systematically be formulated by adopting the EFT methodology (see, for instance, refs. \cite{Petrov,Brivio,Burgess}); the anomalous couplings appear in the form of operators with mass dimension higher than four, and the physics they describe is expected to be suppressed by inverse powers of a cut-off, taken as the energy scale of this new physics. As it shall be discussed later on in this contribution, we adopt another standpoint to generate anomalous couplings. Instead of following the prescriptions of the EFTs, we start off with a non-linear formulation of the weak hypercharge symmetry and describe the $U_Y(1)$ gauge boson Lagrangian by means of the (Abelian) Born-Infeld term \cite{BI_PRSL_34}, similar to the one proposed in the electromagnetic sector \cite{Plebanski,BB_WS_83}. Following this path, after the Higgs field spontaneously break the electroweak symmetry, $4\gamma$, $3\gamma - Z$, $2\gamma-2Z$, $\gamma-3Z$ and $4Z$, tree-level couplings are naturally generated.
With these considerations in mind, and as already expressed above, the purpose of this work is to further elaborate
on the physical content of this rare decay, $Z \to 3 \, \gamma$.
Using results in the literature, we impose a lower bound on the BI-parameter. Moreover, other bounds are also analyzed from the quartic self-coupling $4\gamma$.

At this stage, it is suitable to highlight that the introduction of non-linearity in the hypercharge sector has been also motivated by
recent investigations on magnetic monopoles in the EW scale  \cite{Arunasalam_EPJC_2017,Fabritiis_Helayel}. For more details, we point out
the reviews \cite{Cho_PTRS_A,Mavromatos_IJMPA_2020}.

The anomalous couplings also yield dispersion relations in the presence of an external magnetic field.
We add the prescription of an uniform magnetic background field to the EW BI model such that the
kinetic part of the gauge neutral bosons $Z$-boson and photon is modified emerging mixing
kinetic terms that depends on the ${\bf B}$ and the BI parameter. Thereby, we obtain the dispersion
relations of the photon and $Z$-boson in the presence of this magnetic background.
These results are in agreement with the dispersion relation obtained previously in \cite{MJNevesPRD2021}
for the photon sector.
In this perspective, and given the recent interest in anomalous neutral couplings, it is instructive to further explore the physical consequences presented by this class of couplings. Specifically, in this work we will also focus attention on the impact of these anomalous couplings on physical observables, in particular the static potential between two probe charges, using the gauge-invariant but path-dependent variables formalism, which provides a physically-based alternative to the Wilson loop approach. In fact, our analysis leads to a static potential which is the sum of a Yukawa-type and a linear potential, leading to the confinement of static charges. Incidentally, the above static potential profile is analogous to that encountered for the electroweak sector of the SM with an anomalous triple gauge boson couplings \cite{Gaete_AHEP_2021}. It is also important to observe that similar results have been obtained from different viewpoints. For example, in connection to a theory of antisymmetric tensor fields that results from the condensation of topological defects as a consequence of the Julia-Toulouse mechanism \cite{Gaete2004}, in gluodynamics in curved space-time \cite{Gaete2008}, also in a Higgs-like model \cite{Gaete2009}, for a Podolsky-axionic electrodynamics \cite{Gaete2012}, and for a minimal supersymmetric extension of the Carroll-Field-Jackiw model for electrodynamics with a Chern-Simons-like Lorentz-symmetry violating term \cite{Gaete2013}. The above connections are of interest from the point of view of providing unifications among diverse models as well as exploiting the equivalence in explicit calculations.
Our work is organized as follows: in Section $2$ we describe a new effective model. In other terms, we introduce the non-linear Born-Infeld (BI) theory in the Abelian sector of the EW model $SU_{L}(2) \times U_{Y}(1)$, and we obtain the correspondent anomalous neutral couplings. After that, in Section $3$, we analyze the constraints on the Born-Infeld parameter. In Section $4$, we consider the electroweak Born-Infeld model under a magnetic background. Next, in Section $5$, we compute the interaction energy for a fermion-antifermion pair (static probe charges) in the presence of a $3\gamma-Z$ coupling under an external magnetic field. Finally, some concluding remarks and perspectives are made in Section $6$.
We adopt the Minkowski metric $\eta^{\mu\nu}=\mbox{diag}\left(+1,-1,-1,-1\right)$, and we choose to work with the natural units $\hbar=c=1$ and $4\pi\epsilon_0=1$. In this unit system, the electric, magnetic fields, and force have squared-energy dimension. Moreover, the conversion of Volt/m and Tesla (T) to the natural system is as follows:  $1 \, \mbox{Volt/m}=2.27 \times 10^{-24} \, \mbox{GeV}^2$ and $1 \, \mbox{T} =  6.8 \times 10^{-16} \, \mbox{GeV}^2$, respectively. In this system unit, the conversion for Newton (N) is $1 \, \mbox{eV}^2 = 8.12 \, \times \, 10^{-3} \, \mbox{N}$.

%
%
\section{The description of the model}
\label{sec2}
We start off the description of the model with the complete EW lagrangian
\begin{eqnarray}\label{LEWnl}
{\cal L}_{EW-nl}={\cal L}_{f}+{\cal L}_{Higgs}+{\cal L}_{gauge-nl} \; .
\end{eqnarray}
The fermion and Higgs sectors of the model are unaltered by the non-linearity.
We resume briefly these sectors below. The fermions sector is given by
\begin{eqnarray}
{\cal L}_{f}=i \, \overline{\psi}_{iL} \gamma^\mu D_\mu \psi_{iL} + i \, \overline{\ell}_{iR} \gamma^\mu D_\mu \ell_{iR} \; ,
\end{eqnarray}
where $\gamma^{\mu}$ are the Dirac matrices, and the covariant derivative operator is
\begin{eqnarray}
D_{\mu}=\partial_\mu  + i \, g \, A_\mu^{\;\,a} \, \frac{\sigma^a}{2} + i \, g' \, Y \, B_\mu \; .
\end{eqnarray}
The $Y$ is the hypercharge generator, $g$ and $g^{\prime}$ are the coupling constants of $SU_{L}(2)$
and $U_{Y}(1)$, respectively. The doublets $\psi_{iL} \, (i=1,2,3)$ set left-handed neutrinos/leptons
$L_{i}=\left( \, \nu_{iL} \; \, \ell_{iL} \right)^{t}$ or left-handed quarks $Q_{iL}=\left( \, u_{iL} \; \, d_{iL} \right)^{t}$,
that transform in the fundamental representation of $SU_{L}(2)$, and the right-handed fields $\ell_{iR}$ are $SU_{L}(2)$ singlets,
like in the EW model. The assignments of charges do not change, such that the electric charge of the particle content obeys the
Gell-Mann$–$Nishijima formula $Q_{em}=I_{3L}+Y$, in which $I_{3L}=\sigma^{3}/2$ is the isospin. The Higgs sector is governed by the lagrangian
\begin{eqnarray}
\mathcal{L}_{Higgs} = \vert D_\mu \Phi \vert^2 - \lambda \left( \Phi^\dagger \Phi - \frac{\mu^2}{  2 \lambda} \right)^2 \; ,
\end{eqnarray}
where $\mu^2$ and $\lambda$ are positive and real parameters. The scalar field $\Phi$ is doublet defined in the fundamental representation of
$SU_{L}(2)$.
The modified gauge sector is read below :
\begin{eqnarray}\label{Lgauge}
{\cal L}_{gauge-nl}=-\frac{1}{4} \, F_{\mu\nu}^{\,\,\;\;a}\,F^{\mu\nu a}
-\frac{1}{4} \, B_{\mu\nu}^2+{\cal L}_{nl}({\cal F}_{B} , {\cal G}_{B}) \; , \;\;\;
\end{eqnarray}
where $F_{\mu\nu}^{\;\;\;\,a}=\partial_{\mu}A_{\nu}^{\,\;a}-\partial_{\nu}A_{\mu}^{\,\;a}+g\, \varepsilon^{abc}A_{\mu}^{\,\;b}\,A_{\nu}^{\,\;c} \, (a,b,c=1,2,3)$ is the field strength tensor of $SU_{L}(2)$, $B_{\mu\nu}=\partial_{\mu}B_{\nu}-\partial_{\nu}B_{\mu}$ is the correspondent
one to the  hypercharge group $U_{Y}(1)$. The $A_{\mu}^{\,\;a}$ are the non-Abelian gauge fields of $SU_{L}(2)$, and
$B_{\mu}$ is the Abelian gauge field of $U_{Y}(1)$. We have added the non-linear Lagrangian ${\cal L}_{nl}$ that is function of the Lorentz- and gauge-invariant bilinears ${\cal F}_{B}=-\frac{1}{4} \, B_{\mu\nu}^{2}$ and ${\cal G}_{B}=-\frac{1}{4} \, B_{\mu\nu}\widetilde{B}^{\mu\nu}$,
in which $\widetilde{B}^{\mu\nu}=\epsilon^{\mu\nu\alpha\beta}B_{\alpha\beta}/2$ denotes the dual tensor of $B_{\mu\nu}$. The simplest non-linear
Lagrangian,  which includes anomalous neutral quartic couplings, is given by
\begin{eqnarray}\label{Lnl}
{\cal L}_{nl}({\cal F}_{B} , {\cal G}_{B}) = \xi_{1} \, {\cal F}_{B}^2+ \xi_{2} \, {\cal G}_{B}^2
+ \xi_{3} \, {\cal F}_{B} \, {\cal G}_{B} \; ,
\end{eqnarray}
 with $\xi_{1}$, $\xi_{2}$ and $\xi_{3}$ being real parameters, but notice that $\xi_{3} \neq 0$ implies into the parity  symmetry breaking.
In this paper, we consider the effective interactions generated by the Born-Infeld Lagrangian
\begin{eqnarray}\label{LBI}
{\cal L}_{BI} &=& \beta^{2} \left[ \, 1-\sqrt{1-2\frac{{\cal F}_{B}}{\beta^2} - \frac{{\cal G}_{B}^2}{\beta^4} } \, \right]
\nonumber \\
&\simeq & {\cal F}_{B}+\frac{1}{2\beta^2} \left( \, {\cal F}_{B}^2+{\cal G}_{B}^{2} \, \right) \; ,
\end{eqnarray}
with the real parameter $\beta \gg |{\cal F}_{B}|^{1/2},|{\cal G}_{B}|^{1/2}$. Notice that the usual
kinetic term is recovered in the limit $\beta \rightarrow \infty$. Comparing (\ref{LBI}) and (\ref{Lnl}),
we obtain $\xi_{1}=\xi_{2}=1/(2\beta^2)$ and $\xi_{3}=0$. Therefore, it is immediate that the Lagrangian (\ref{LEWnl})
keeps the gauge invariance $SU_{L}(2) \times U_{Y}(1)$ in the EW gauge sector.
The SSB mechanism takes place in the same way as it occurs in the SM.
The potential has its minimal at the vacuum expectation value $v=\sqrt{\mu^2/(2\lambda)}=246$ GeV,
and we expand the scalar field around the $v$ - scale choosing the unitary gauge
$\Phi=\left[\,\, 0 \;\;\; (v+h)/\sqrt{2} \,\,\right]^{t}$. After the SSB, the charged gauge bosons
$W^{\pm}$ acquire mass of $m_{W}=80$ GeV, and the kinetic neutral sector of the gauge bosons is given by
\begin{eqnarray}\label{LKgaugenl}
{\cal L}^{(kin)}_{gauge-nl}=-\frac{1}{4} \, (\partial_{\mu}A_{\nu}^{\,\;3}-\partial_{\nu}A_{\mu}^{\,\;3})^2
-\frac{1}{4} \, B_{\mu\nu}^2
\nonumber \\
+\frac{v^2}{8}(g\, A_{\mu}^{3}-g^{\prime}B_{\mu})^{2}
+\frac{1}{2\beta^2} \left( \, {\cal F}_{B}^2+{\cal G}_{B}^{2} \, \right) \; .
\end{eqnarray}
%
%
%
The physical eigenstates of the neutral gauge bosons are obtained by the well-known Weinberg angle rotation:
\begin{subequations}
\begin{eqnarray}
A_{\mu}^{\;\,3} &=& \cos\theta_{W} \, Z_{\mu} + \sin\theta_{W} \, A_{\mu} \; ,
\label{Amu3}
\\
B_{\mu} &=& -\sin\theta_{W} \, Z_{\mu} + \cos\theta_{W} \, A_{\mu} \; ,
\label{Bmu}
\end{eqnarray}
\end{subequations}
where $\theta_W$ is the Weinberg angle, for which $\sin^{2}\theta_{W} = 0.23$.
The physical gauge field $Z_{\mu}$ sets the $Z$-gauge boson with mass of $m_{Z}=91$ GeV,
and $A_{\mu}$ is identified as the massless photon field. Using the transformations (\ref{Amu3}) and (\ref{Bmu})
in (\ref{LKgaugenl}), the neutral sector can be written as
\begin{eqnarray}
{\cal L}^{(kin)}_{gauge-nl}={\cal L}^{(kin)}_{gauge}+{\cal L}_{BI-nl} \; .
\end{eqnarray}
where the kinetic sector is
\begin{eqnarray}
{\cal L}^{(kin)}_{gauge}=-\frac{1}{4} \, F_{\mu\nu}^{2}-\frac{1}{4} \, Z_{\mu\nu}^{2}+\frac{1}{2} \, m_{Z}^2 \, Z_{\mu}^2 \; .
\end{eqnarray}
The effective non-linear Born-Infeld sector contributions formulated in the physical basis are given by
\begin{eqnarray}\label{LBInl}
{\cal L}_{BI-nl}={\cal L}^{4\gamma}+{\cal L}^{3\gamma-Z}+{\cal L}^{2\gamma-2Z}+{\cal L}^{\gamma-3Z}+{\cal L}^{4Z} \, ,
\;\;\;\;\;
\end{eqnarray}
where the anomalous neutral couplings read as it follows below:
\begin{subequations}
\begin{eqnarray}
{\cal L}^{4\gamma} &=& \frac{ \cos^4 \theta_W }{32\beta^2} \left[ \, \left(F_{\mu\nu}^2\right)^2
+ \left( F_{\mu\nu}\widetilde{F}^{\mu\nu} \right)^2 \, \right] \; ,
\label{L4gamma}
\\
{\cal L}^{3\gamma-Z} &=& -\frac{\sin(2\theta_{W})}{16\beta^2} \, \cos^2\theta_{W}
\left[ \, \left(F_{\mu\nu}^2\right) \left(F_{\alpha\beta}Z^{\alpha\beta}\right)+
\right.
\nonumber \\
&&
\left.
\hspace{-0.5cm}
+\left(F_{\mu\nu}\widetilde{F}^{\mu\nu} \right)
\left( F_{\alpha\beta}\widetilde{Z}^{\alpha\beta} \right) \, \right] \; ,
\label{L3gammaZ}
\\
{\cal L}^{2\gamma-2Z} \!&=&\! \frac{\sin^2(2\theta_{W})}{32\beta^2}
\left[ \, \frac{1}{2} \, \left(F_{\mu\nu}^2\right) \, \left(Z_{\alpha\beta}^2\right)
+\left(F_{\mu\nu}Z^{\mu\nu} \right)^2 \right. +
\nonumber \\
&&
\left.
\hspace{-0.5cm}
 + \frac{1}{2} \left( F_{\mu\nu}\widetilde{F}^{\mu\nu} \right) \left(
Z_{\alpha \beta} \widetilde{Z}^{\alpha \beta} \right)
+\left( F_{\mu\nu}\widetilde{Z}^{\mu\nu} \right)^2 \, \right] \, , \;\;\;
\label{L2gamma2Z}
\\
{\cal L}^{\gamma-3Z} &=& -\frac{\sin(2\theta_{W})}{16\beta^2} \, \sin^2\theta_{W}
\left[ \, \left(Z_{\mu\nu}^2\right) \left(F_{\alpha\beta}Z^{\alpha\beta}\right)+
\right.
\nonumber \\
&&
\left.
\hspace{-0.5cm}
+\left(F_{\mu\nu}\widetilde{Z}^{\mu\nu} \right)
\left( Z_{\alpha\beta}\widetilde{Z}^{\alpha\beta} \right) \, \right] \; ,
\label{Lgamma3Z}
\\
{\cal L}^{4Z} &=& \frac{ \sin^4 \theta_W }{32\beta^2} \left[ \, \left(Z_{\mu\nu}^2\right)^2
+ \left( Z_{\mu\nu}\widetilde{Z}^{\mu\nu} \right)^2 \, \right] \; ,
\label{L4Z}
\end{eqnarray}
\end{subequations}
and $Z_{\mu\nu}=\partial_{\mu}Z_{\nu}-\partial_{\nu}Z_{\mu}$, $F_{\mu\nu}=\partial_{\mu}A_{\nu}-\partial_{\nu}A_{\mu}$
denote the $Z$- and photon field strength tensors. The $\widetilde{Z}^{\mu\nu}=\epsilon^{\mu\nu\alpha\beta}Z_{\alpha\beta}/2$ and $\widetilde{F}^{\mu\nu}=\epsilon^{\mu\nu\alpha\beta}F_{\alpha\beta}/2$ are the dual field strength tensors that satisfy the Bianchi
identities $\partial_{\mu}\widetilde{Z}^{\mu\nu}=0$ and $\partial_{\mu}\widetilde{F}^{\mu\nu}=0$, respectively. All these quartic couplings
are not contained in the SM, and go to zero in the limit $\beta \rightarrow \infty$. They give rise to new effects,
as the mixed interactions of the photon with the $Z$-gauge boson, involving $3\gamma-Z$, $2\gamma-2Z$ and $\gamma-3Z$, and the quartic couplings of
$A^{\mu}$, and of $Z^{\mu}$. In what follows, we investigate the consequences of the coupling $3\gamma-Z$ to bound the BI $\beta$-parameter.
\section{Constraining the Born-Infeld parameter}
\label{sec3}
In this Section, let us initially analyze the decay $Z \rightarrow 3 \gamma$ that emerges from the new coupling $3\gamma - Z$ in eq.
(\ref{L3gammaZ}).
This anomalous neutral coupling has been a subject of intense research in the literature.
The phenomenological Lagrangian is described by
\begin{equation} \label{L_phe}
{\cal L}^{3\gamma-Z}_{ph} = \zeta_1 \left(F_{\mu\nu}^2\right) \left(F_{\alpha\beta}Z^{\alpha\beta}\right)
+ \zeta_2 \left(F_{\mu\nu}\widetilde{F}^{\mu\nu} \right)
\left( F_{\alpha\beta}\widetilde{Z}^{\alpha\beta} \right) \; ,
\end{equation}
with $\zeta_1$ and $\zeta_2$ being two parameters.
%

%

We emphasize that similar interactions can be gene-rated in an effective theory approach to the SM, by taking into account loop corrections. 
For a detailed review, we highlight ref. \cite{Green_RMP}. However, we adopt another viewpoint in which the anomalous neutral couplings appear as contributions of new physics. Specifically, in our contribution, we take the effects of new physics as a non-linear Abelian physics given by the action associated with the $U_{Y}(1)$. In this, we differ from the conventional effective field theory treatment. For this reason, we shall neglect the quantum corrections of the SM and use some phenomenological investigations to get lower bounds for the BI parameter. At this stage, it should be mentioned that the anomalous couplings \eqref{L_phe} were also obtained in other proposals beyond SM, involving new neutral particles (spin-0 and spin-2 excitations) \cite{Baldenegro_JHEP_2017}.

For the non-linear BI extension in the hypercharge sector (\ref{LBI}), we arrive at
\begin{equation} \label{zeta_def}
\zeta_1 \equiv \zeta_2 = -\frac{\sin(2\theta_{W})}{16\beta^2} \cos^2\theta_{W} \, .
\end{equation}
Here, we take advantage of the phenomenological results to constraint the $\beta-$parameter. According to ref.
\cite{Baldenegro_JHEP_2017}, the $3\gamma-Z$ coupling describes the $Z$ decay into three photons $Z \rightarrow 3\gamma$. With the aforementioned coupling constants (\ref{zeta_def}), we obtain the following decay width at the tree level
\begin{eqnarray}
\Gamma(Z \rightarrow 3\gamma)=\frac{3m_{Z}}{2160 \pi^3} \, \sin^2\theta_{W}\cos^{6}\theta_{W}
\left( \frac{m_{Z}^2}{4\beta} \right)^4 \, , \;\;
\end{eqnarray}
and using that $m_{Z}=91$ GeV and $\sin^{2}\theta_{W}=0.23$, we end up with the numerical result in terms of the
$\beta$-parameter
\begin{eqnarray}
\Gamma(Z \rightarrow 3\gamma)= 0.42845 \, \mbox{MeV} \left( \frac{2070.25 \, \mbox{GeV}^2 }{\beta\,[\mbox{GeV}^2]} \right)^4 \, .
\end{eqnarray}
The branch ratio for this process is read below
\begin{eqnarray}
\mbox{Br}(Z \rightarrow 3\gamma)=\frac{\Gamma(Z \rightarrow 3\gamma)}{\Gamma_{Z}+\Gamma(Z \rightarrow 3\gamma)} \; ,
\end{eqnarray}
where $\Gamma_{Z}=2.4952 \, \pm \, 0.0023 \, \mbox{GeV}$ is the full $Z$-decay width in the SM \cite{PDG2018}. Using the upper bound for the branch ratio
$\mbox{Br}(Z \rightarrow 3\,\gamma) < 2.2 \times 10^{-6}$ \cite{AtlasEPJC2016}, we obtain the lower bound
%
\begin{eqnarray}\label{sqrtbetabound}
\sqrt{\beta} > 78.62 \, \mbox{GeV} \; .
\end{eqnarray}
In the work of ref. \cite{PdeFabrittis}, the reader may find more details to get a similar bound. Furthermore, this result exhibits the same order as the one obtained in ref. \cite{Ellis_PRL_2017} $(\sqrt{\beta} \gtrsim 100 \, \, \mbox{GeV}) $, where the authors analyzed the measurement of light-by-light scattering.
%
%
%
%
%
%
%
%

The attainment of the lower-bound expressed in the eq. (\ref{sqrtbetabound}) above on the Born-Infeld $\beta$-parameter, is – we highlight –  one of the main motivations of the present Section. On the other hand, in view of the Landau-Yang theorem (LY) \cite{Landau,Yang50}, which states that a massive vector (or pseudo-vector) boson (in our specific case, the $Z$-boson) cannot decay into two on-shell photons on the basis of first principles -  by virtue of both Lorentz and gauge invariance and the symmetry of the two-photon final state - the anomalous couplings of $Z$ to three photons given in the Lagrangian (\ref{L3gammaZ}) opens up a natural path to constrain the Born-Infeld parameter by adopting the results reported in 2016 by ATLAS Collaboration \cite{AtlasEPJC2016}.  Nevertheless, in this context, we could call into question the possibility to consider the $Z$ decay into two photons in presence of an external magnetic field. This magnetic field may naturally be brought about if we split the field strength, $F_{\mu\nu}$, appearing in (\ref{L3gammaZ}) into a background field, $F_{0\mu\nu}$, (actually, the external magnetic field, $F_{0ij} \sim \epsilon_{ijk} B_k$) and a perturbation that describes the photon field, $f_{\mu\nu}$ . In so doing, $Z\gamma\gamma-$vertices are generated out of the Lagrangian (\ref{L3gammaZ}), from which the dematerialization of $Z$ into a photon pair can be studied. However, the emergent photons do not obey the usual Maxwell equations, for they propagate in a magnetic background and they have non-trivial mixings with the $Z$-field. Therefore, they obey non-trivial modified dispersion relations. In addition,  the relation between the polarization planes of the photons are affected by the external magnetic field and the mixing with the $Z$-field. This may be readily seen from eq. (\ref{fieldeqfmunu}) of Section IV, which describes the propagation of the emergent photons in the magnetic background and displays two mixing terms with the $Z$-field . The photon's on-shell profile has therefore to be carefully studied to account for the full effect of the external field. The re-evaluation of the Landau-Yang disintegration theorem in this scenario appears as an important issue to be pursued, and we shall be carrying out this investigation to report on it in a forthcoming work. Incidentally, let us point out the interesting work by Ivanov and coworkers where they discuss if twisted photons might overcome the restrictions imposed by the Landau-Yang theorem, so that the decay could take place \cite{Ivanov}.

Other bounds for $\beta$-parameter are associated with effects from ${\cal L}^{4\gamma}$ in eq. (\ref{L4gamma}).
This quartic photon self-coupling can be written as
\begin{eqnarray}
{\cal L}^{4\gamma} &=& \zeta_{3} \, \left(F_{\mu\nu}^2\right)^2
+ \zeta_{4} \, F_{\mu\nu}F^{\mu\rho}F_{\rho\sigma}F^{\sigma\nu} \; ,
\end{eqnarray}
where $\zeta_{3}$ and $\zeta_{4}$ are defined by
\begin{eqnarray}
\zeta_{3}= - \frac{ \cos^4 \theta_W }{ 32\beta^2}
\; \; , \; \;
\zeta_{4} = - \frac{ \cos^4 \theta_W }{8\beta^2} \; .
\end{eqnarray}
Using the bounds $|\zeta_{3}| < 2.88 \times 10^{-13}\, \mbox{GeV}^{-4}$ and $|\zeta_{4}| < 6.02
\times 10^{-13}\, \mbox{GeV}^{-4}$ of recent investigations from the CMS and TOTEM Collaborations \cite{CMSTOTEM},
we obtain stronger limits, respectively :
\begin{subequations}
\begin{eqnarray}
\sqrt{\beta} &>& 502.84 \, \mbox{GeV} \; ,
\label{beta1}
\\
\sqrt{\beta} &>& 591.42 \, \mbox{GeV} \; .
\label{beta2}
\end{eqnarray}
\end{subequations}
In view of the bounds (\ref{beta1}) and (\ref{beta2}), we notice that they are compatible with the previous bounds attained from refs. \cite{PdeFabrittis} and \cite{Ellis_PRL_2017}. However, we point out that the difference in the estimates of $\beta$-parameter by using the approaches of effective field theory and the non-linear Abelian description of the hypercharge sector should still be confirmed by the forthcoming results of the LHC in its third run.



In addition to these lower bounds for the $\beta$-parameter (expressed in magnetic field units, $\beta \gtrsim  \, 10^{19} \, \mbox{T}$), we would like to make a few considerations on the dispersion relations that emerge from the generalized Born-Infeld action of the work in ref. \cite{GaeteEPJC}. The correspondent dispersion relations and the kinematics of Compton scattering have been discussed in ref. \cite{MJNevesPRD2021},  whenever an external uniform magnetostatic field, ${\bf B}$, is present.
%


If we consider the traditional Born-Infeld action, the effect of the non-linearity combined with the influence of the external magnetic field yield a redshift of the waves $(\omega_{BI} < \omega_{M})$; its expression is as follows below:
\begin{eqnarray}\label{Deltaomega}
\Delta\omega=|\omega_{BI}-\omega_{M}|=\frac{2\pi}{\lambda}\left[ \, 1-\sqrt{\frac{\beta^2+{\bf B}^2\cos^2\alpha}{\beta^2+{\bf B}^2}} \, \right]
\, . \;\;\;\;\;
\end{eqnarray}
The $\omega_{M}$ and $\omega_{BI}$ are, respectively, the frequencies associated with the (same) wavelength, $\lambda$, in the Maxwellian and Born-Infeld cases, whereas $\alpha$ is the angle between the wave vector, ${\bf k}$, and the external magnetic field. Notice that there is no redshift whenever the wave propagates along the ${\bf B}$-field direction; on the other hand, the maximal value for the redshift occurs in the case the wave propagates along a direction orthogonal to ${\bf B}$.
%


By using $\beta \sim 10^{19} \, \mbox{T}$ and an external magnetic field typical of the era between the Electroweak and QCD phase transitions, $|{\bf B}| \sim 10^{17} \, \mbox{T}$, and considering the situation of maximal redshift $(\cos \alpha = 0)$ , the latter is estimated to be of the order of $9.39$ Hz in the radio-frequency region, and $9.39 \times 10^{12}$ Hz in the X-ray domain. Under these conditions, the ratio between (\ref{Deltaomega}) and $\omega_{M}$ corresponds to $\Delta\omega/\omega_{M}\sim 0.005 \%$.
The dispersion relations worked out in Section V of ref. \cite{MJNevesPRD2021} for the generalized Born-Infeld action might potentially exhibit dichroism for the power $p$ in the range $0 < p < 0.5$ . For that, the inequality below should be satisfied :
\begin{eqnarray}
\beta^2+(2p-1) \, {\bf B}^2+2 \, (1-p) \, {\bf B}^2\cos^{2}\alpha < 0 \; .
\end{eqnarray}
However, in view of the bound $\beta \gtrsim  \, 10^{19} \, \mbox{T}$ and considering that magnetic fields with magnitude above $\beta$ destabilize the Higgs vacuum and, therefore, they are excluded, the relation above is never fulfilled for $0 < p < 0.5$. We then conclude that, with these constraints for the $\beta$-parameter, our generalized Born-Infeld action exhibits birefringence for $p$ different from $0.5$ (the original Born-Infeld does not exhibit birefringence),  but dichroism never takes place over the whole range $(0<p<1)$.

\section{The electroweak BI model in presence of a magnetic background}
We add the prescription of the electroweak BI model in a uniform magnetic background by expanding the
$A^{\mu}$ potential as $A_{\mu}=a_{\mu}+A_{0\mu}$, where $a^{\mu}$ stands for the photon gauge field, and
$A_{0}^{\;\;\mu}=(0,A_{0}^{\;\;\,i})$ is the background potential, with $A_{0}^{\;\;i}=\epsilon^{ijk}x^{j}B^{k}/2$,
and $i,j,k=1,2,3$. The expansion in the field strength tensor is $F_{\mu\nu}=f_{\mu\nu}+F_{0\mu\nu}$, where
$f^{\mu\nu}=\partial^{\mu}a^{\nu}-\partial^{\nu}a^{\mu}$ corresponds to the photon field strength tensor,
and $F_{0}^{\;\;\mu\nu}=(0,-\epsilon^{ijk}B^{k})$ denotes the magnetic background tensor.
The dual tensor for this magnetic background is $\tilde{F}_{0}^{\;\;\mu\nu}=(-B^{i},0)$.

The kinetic sector of the boson $Z$ with the photon in the $F_{0\mu\nu}$ background is
\begin{eqnarray}\label{intZphoton}
{\cal L}_{kin}^{\gamma-Z} &=& -\frac{1}{4} \, f_{\mu\nu}^2-\frac{1}{4} \, Z_{\mu\nu}^2+\frac{1}{2} \, m_{Z}^2 \, Z_{\mu}^{2}+
\nonumber \\
&&
\hspace{-1cm}
+\frac{\cos^4\theta_{W}}{16\beta^2}\left[\, f_{\mu\nu}^2\,F_{0\alpha\beta}^2+2\,(f_{\mu\nu}F_{0}^{\;\,\mu\nu})^{2}
\right.
\nonumber \\
&&
\hspace{-1cm}
\left.
+(f_{\mu\nu}\tilde{f}^{\mu\nu})(F_{0\alpha\beta}\tilde{F}_{0}^{\;\,\alpha\beta}) + 2 \, (f_{\mu\nu}\tilde{F}_{0}^{\;\,\mu\nu})^2 \,  \right]
\nonumber \\
&&
\hspace{-1cm}
+\frac{\sin^2(2\theta_{W})}{32\beta^2}
\left[ \, \frac{1}{2} \, \left(F_{0\mu\nu}^2\right) \, \left(Z_{\alpha\beta}^2\right)
+\left(F_{0\mu\nu}Z^{\mu\nu} \right)^2
\right.
\nonumber \\
&&
\left.
\hspace{-1cm}
 + \frac{1}{2} \left( F_{0\mu\nu}\widetilde{F}_{0}^{\;\,\mu\nu} \right) \left(
Z_{\alpha \beta} \widetilde{Z}^{\alpha \beta} \right)
+\left( F_{0\mu\nu}\widetilde{Z}^{\mu\nu} \right)^2 \, \right]
\nonumber \\
&&
\hspace{-1cm}
-\frac{\sin(2\theta_{W})}{16\beta^2} \, \cos^2\theta_{W}\left[ \, F_{0\mu\nu}^2 \, f_{\alpha\beta}Z^{\alpha\beta}+
\right.
\nonumber \\
&&
\left.
\hspace{-1cm}
+2\,(f_{\mu\nu}F_{0}^{\;\,\mu\nu})(Z_{\alpha\beta}F_{0}^{\;\,\alpha\beta})
+2\,(f_{\mu\nu}\tilde{F}_{0}^{\;\,\mu\nu})(Z_{\alpha\beta}\tilde{F}_{0}^{\;\,\alpha\beta})
\right.
\nonumber \\
&&
\left.
\hspace{-1cm}
+(F_{0\mu\nu}\tilde{F}_{0}^{\;\,\mu\nu})(f_{\alpha\beta}\tilde{Z}^{\alpha\beta}) \,
\right]
\, .
\;\;\;\;\;\;
\end{eqnarray}
This prescription yields kinetic mixing of the $Z$-gauge bosons with the photon
$a^{\mu}$, as for example, into the form
\begin{eqnarray}
{\cal L}_{mix}^{\gamma-Z} = -\sin\theta_{W}\cos^3\theta_{W} \left(\frac{{\bf B}^2}{2\beta^2}\right) \, \frac{1}{2} \,  f_{\mu\nu}\,Z^{\mu\nu}
\nonumber \\
-\sin\theta_{W}\cos^3\theta_{W} \left(\epsilon^{ijk}\epsilon^{mnl} \, \frac{B^{k}B^{l}}{2\beta^2}\right) \frac{1}{2} \, f_{ij}\,Z_{mn}
\nonumber \\
-\sin\theta_{W}\cos^3\theta_{W} \left(\frac{B^{i}B^{j}}{\beta^2}\right) f_{0i} \, Z_{0j}
 \; ,
\end{eqnarray}
whose the mixing parameter has the magnitude like $\sin\theta_{W}\cos^3\theta_{W} \, {\bf B}^2/\beta^2$.
Using the Schwinger's critical magnetic field $B_{c}\simeq 10^{-6} \, \mbox{GeV}^2$, and the
parameter $\beta \sim 10^{4} \, \mbox{GeV}^2$, the kinetic mixing is estimated at ${\bf B}^2/\beta^2\sim 10^{-20}$.
For the case of neutron stars, the magnetic field is $|{\bf B}| \simeq 10^{-5} \, \mbox{GeV}^2$,
that reproduces a kinetic mixing at order of ${\bf B}^2/\beta^2\sim 10^{-17}$.
Other example of the SM, is the magnetic field in QCD phase transition of $|{\bf B}| \simeq 10 \, \mbox{GeV}^2$
in which the kinetic mixing is ${\bf B}^2/\beta^2\sim 10^{-8}$. Thus, following these examples, the kinetic
mixing has a small contribution into the field propagation.
The action principle leads to the $Z$-boson and photon field equations :
\begin{subequations}
\begin{eqnarray}\label{fieldeqfZmunu}
&&
\left[ 1+\frac{\cos^4\theta_{W}}{\beta^2}\!\left(-\frac{1}{4}F_{0\alpha\beta}^2\right) \right]\partial^{\mu}f_{\mu\nu}
\nonumber \\
&&
-\frac{\cos^4\theta_{W}}{2\beta^2} \, R_{\mu\nu\alpha\beta} \, \partial^{\mu}f^{\alpha\beta}
\nonumber \\
&&
-\frac{\sin\theta_{W}\cos^3\theta_{W}}{\beta^2} \left(-\frac{1}{4}F_{0\alpha\beta}^2 \right) \partial^{\mu}Z_{\mu\nu}
\nonumber \\
&&
+\frac{\sin\theta_{W}\cos^3\theta_{W}}{2\beta^2} \, R_{\mu\nu\alpha\beta} \, \partial^{\mu}Z^{\alpha\beta}=0 \; ,
\label{fieldeqfmunu}
\\
&&
\left[ 1+\frac{\sin^2(2\theta_{W})}{4\beta^2}\left(-\frac{1}{4}F_{0\alpha\beta}^2\right) \right]\partial^{\mu}Z_{\mu\nu}+m_{Z}^2\,Z_{\nu}
\nonumber \\
&&
-\frac{\sin^2(2\theta_{W})}{8\beta^2} \, R_{\mu\nu\alpha\beta} \, \partial^{\mu}Z^{\alpha\beta}
\nonumber \\
&&
-\frac{\sin\theta_{W}\cos^3\theta_{W}}{\beta^2} \left(-\frac{1}{4}F_{0\alpha\beta}^2\right) \partial^{\mu}f_{\mu\nu}
\nonumber \\
&&
-\frac{\sin\theta_{W}\cos^3\theta_{W}}{2\beta^2} \, R_{\mu\nu\alpha\beta}  \, \partial^{\mu}f^{\alpha\beta}=0 \; ,
\label{fieldeqZmunu}
\end{eqnarray}
\end{subequations}
where we have defined the background tensor $R_{\mu\nu\alpha\beta}=F_{0\mu\nu}F_{0\alpha\beta}+\tilde{F}_{0\mu\nu}\tilde{F}_{0\alpha\beta}$,
that is antisymmetric exchanging $\mu \leftrightarrow \nu$ or $\alpha \leftrightarrow \beta$, and symmetric
for $\mu\nu \leftrightarrow \alpha\beta$. This $Z$ equation implies into the subsidiary condition $\partial_{\mu}Z^{\mu}=0$, and the photon gauge symmetry allow us to choose the Lorenz gauge $\partial_{\mu}a^{\mu}=0$. Under these conditions, the previous field equations can be written in terms of the
potentials $Z^{\mu}$ and $a^{\mu}$, respectively,
%
%
\begin{subequations}
\begin{eqnarray}\label{Eqspotentials}
&&
\left[ 1+\frac{\cos^4\theta_{W}}{\beta^2}\left(-\frac{1}{4}F_{0\alpha\beta}^2\right) \right]\Box a_{\nu}
\nonumber \\
&&
-\frac{\cos^4\theta_{W}}{\beta^2}\, R_{\mu\nu\alpha\beta} \, \partial^{\mu}\partial^{\alpha}a^{\beta}
\nonumber \\
&&
-\frac{\sin\theta_{W}\cos^3\theta_{W}}{\beta^2} \left(-\frac{1}{4}F_{0\alpha\beta}^2 \right) \Box Z_{\nu}
\nonumber \\
&&
+\frac{\sin\theta_{W}\cos^3\theta_{W}}{\beta^2} \, R_{\mu\nu\alpha\beta} \, \partial^{\mu}\partial^{\alpha}Z^{\beta}=0 \; ,
\\
&& \label{Eqspotentials2}
\left[ 1+\frac{\sin^2(2\theta_{W})}{4\beta^2}\left(-\frac{1}{4}F_{0\alpha\beta}^2\right) \right]\Box Z_{\nu}+m_{Z}^2\,Z_{\nu}
\nonumber \\
&&
-\frac{\sin^2(2\theta_{W})}{4\beta^2} \, R_{\mu\nu\alpha\beta} \,
\partial^{\mu}\partial^{\alpha}Z^{\beta}
\nonumber \\
&&
-\frac{\sin\theta_{W}\cos^3\theta_{W}}{\beta^2} \left(-\frac{1}{4}F_{0\alpha\beta}^2\right) \Box a_{\nu}
\nonumber \\
&&
-\frac{\sin\theta_{W}\cos^3\theta_{W}}{\beta^2} \, R_{\mu\nu\alpha\beta} \, \partial^{\mu}\partial^{\alpha}a^{\beta}=0 \; .
\end{eqnarray}
\end{subequations}
Using the plane wave solutions $a^{\mu}(x)=a_{0}^{\,\;\mu} \, e^{i \, k \cdot x}$
and $Z^{\mu}(x)=z_{0}^{\,\;\mu} \, e^{i \, k \cdot x}$, in which $a_{0}^{\,\;\mu}$ and $z_{0}^{\,\;\mu}$ are uniform amplitudes,
we can write the eqs. (\ref{Eqspotentials}) and (\ref{Eqspotentials2}) into the matrix form
\begin{eqnarray}\label{Mijzj}
\left[
\begin{array}{cc}
A_{\mu\nu} & B_{\mu\nu}
\\
\\
C_{\mu\nu} & D_{\mu\nu} \\
  \end{array}
\right]
\left(
\begin{array}{c}
a_{0}^{\;\,\nu}
\\
\\
z_{0}^{\;\,\nu} \\
\end{array}
\right)
=0 \; ,
\end{eqnarray}
where matrix elements are defined by
%
\begin{subequations}
\begin{eqnarray}
A_{\mu\nu} &=& \left[ 1+\frac{\cos^4\theta_{W}}{\beta^2}\left(-\frac{1}{4}F_{0\alpha\beta}^2\right) \right](-k^2) \, \eta_{\mu\nu}
\nonumber \\
&&
\hspace{-0.5cm}
+\frac{\cos^4\theta_{W}}{\beta^2}\, R_{\mu\alpha\nu\beta} \,
k^{\alpha}\,k^{\beta} \; ,
\\
B_{\mu\nu} &=& \frac{\sin\theta_{W}\cos^3\theta_{W}}{\beta^2}\left(-\frac{1}{4}F_{0\alpha\beta}^2\right) k^2 \, \eta_{\mu\nu}
\nonumber \\
&&
\hspace{-0.5cm}
-\frac{\sin\theta_{W}\cos^3\theta_{W}}{\beta^2}\, R_{\mu\alpha\nu\beta} \,
k^{\alpha}\,k^{\beta} \; ,
\\
C_{\mu\nu} &=& \frac{\sin\theta_{W}\cos^3\theta_{W}}{\beta^2}\left(-\frac{1}{4}F_{0\alpha\beta}^2\right) k^2 \, \eta_{\mu\nu}
\nonumber \\
&&
\hspace{-0.5cm}
+\frac{\sin\theta_{W}\cos^3\theta_{W}}{\beta^2}\, R_{\mu\alpha\nu\beta} \,
k^{\alpha}\,k^{\beta} \; ,
\\
D_{\mu\nu} &=& \left[ 1+\frac{\sin^2(2\theta_{W})}{4\beta^2}\left(-\frac{1}{4}F_{0\alpha\beta}^2\right) \right](-k^2) \, \eta_{\mu\nu}
\nonumber \\
&&
\hspace{-0.5cm}
+ \, m_{Z}^2 \, \eta_{\mu\nu}
+\frac{\sin^2(2\theta_{W})}{4\beta^2}\, R_{\mu\alpha\nu\beta} \,
k^{\alpha}\,k^{\beta} \; .
\end{eqnarray}
\end{subequations}
%
%
%
The non-trivial solution of (\ref{Mijzj}) implies that the determinant of the matrix is null.
The possible frequency solutions are : $\omega_{i}^{\pm}=\pm \, \omega_{i}({\bf k}) \, (i=1,2,3,4)$, where
$\omega_{1}({\bf k})=|{\bf k}|$ and $\omega_{2}({\bf k})=\sqrt{ {\bf k}^2+m_{Z}^2 }$ are the known dispersion relations
of the photon and $Z$ gauge boson, respectively. Taking into account the approximation $\beta \gg |{\bf B}| $, the two new solutions read below :

\begin{subequations}
\begin{eqnarray}
\omega_{3}({\bf k}) & \simeq & |{\bf k}| \, \left[ \, 1 - ({\bf B}\times\hat{{\bf k}})^{2} \, \frac{\cos^4\theta_{W}}{2\beta^2} \, \right]
\; ,
\label{omega3ap}
\\
\omega_{4}({\bf k}) & \simeq &
\sqrt{{\bf k}^2+m_{Z}^2} + \frac{{\bf B}^2\,m_{Z}^2 - ({\bf B}\times{\bf k})^{2}}{ \sqrt{{\bf k}^2+m_{Z}^2} } \, \frac{\sin^2(2\theta_{W})}{8\beta^2}
\; .
\nonumber \\
\label{omega4ap}
\end{eqnarray}
\end{subequations}
%
We recover the usual results for frequencies of the photon and $Z$ in the limit $\beta \rightarrow \infty$. The small correction yields the positive frequency (\ref{omega4ap}) if the $\beta$-parameter satisfies the condition
%
\begin{eqnarray}
\beta > 0.29 \, \sqrt{ \, \frac{({\bf B}\times{\bf k})^2-{\bf B}^2\,m_{Z}^2}{{\bf k}^2+m_{Z}^2} \, } \; ,
\label{beta4}
\end{eqnarray}
%
with the wave vector constraint by $|\hat{{\bf B}}\times {\bf k}| > 91$ GeV. For the wave propagation direction perpendicular
to magnetic field direction, this condition constraints the wavelength of the $Z$-wave by $\lambda < 0.070 \, \mbox{GeV}^{-1}\simeq 1.38 \times 10^{-17} \, \mbox{m}$.

Within this approximation, it is interesting to highlight that the frequency (\ref{omega3ap}) converges to non-trivial dispersion relation in the photon sector due to the Euler-Heisenberg-like coupling, namely, ${\cal L}^{4\gamma}$ in (\ref{L4gamma}). Indeed, by using the procedure described in ref. \cite{MJNevesPRD2021} for this quartic self-coupling in the presence of magnetic background, one arrives at the same result.

\section{Interaction energy for $3\gamma-Z$ under an uniform magnetic field}
As already stated, we now turn our attention to examine here the effects of these new anomalous couplings on a physical observable.
To do this, we will work out the static potential for the model under consideration by using the gauge-invariant but path-dependent variables formalism, along the lines of Ref. \cite{Gaete97,Gaete_AHEP_2021}. To be more precise, we shall compute the expectation value of the energy operator $H$ in the physical state $\left| \Phi  \right\rangle$, which we denote by $\langle H\rangle _\Phi$. For simplicity, in the present analysis we shall consider a gauge theory which describes the interaction between the familiar massless $U(1)_{em}$ photon with the massive vector $Z$-field via the new coupling ${\cal L}^{3\gamma-Z}$ in (\ref{L3gammaZ}). In such a case, the Lagrangian density reads
\begin{eqnarray}\label{energy-05}
{\cal L} &=& - \, \frac{1}{4} \, F_{\mu \nu }^2 - \frac{1}{4} \, Z_{\mu \nu }^2 + \frac{1}{2} \, m_Z^2 \, Z_{\mu}^{2} +
\nonumber \\
&&
\hspace{-0.5cm}
-\frac{\sin \theta_W }{8\beta ^2} \cos^3\theta_W \left[ {\left( {F_{\mu \nu }^2} \right)\left( {{F_{\alpha \beta }}{Z^{\alpha \beta }}} \right)} \right]
\nonumber \\
&&
\hspace{-0.5cm}
-\frac{\sin\theta _W}{8\beta^2} \cos^3\theta _W \left[ {\left( {{F_{\mu \nu }}{{\tilde F}^{\mu \nu }}} \right)\left( {{F_{\alpha \beta }}{{\tilde Z}^{\alpha \beta }}} \right)} \right] \; ,
\hspace{0.5cm}
\end{eqnarray}
%
%
Next, if we consider the model in the limit of a very heavy $Z$-field and we are bound to energies much below $m_{Z}$, we are allowed to integrate over $Z_{\mu}$ and to speak about an effective model for the $A_{\mu}$-field. This can be readily accomplished by means of the path integral formulation of the generating functional associated with the eq. (\ref{energy-05}), where the $Z_{\mu}$-field appears at most quadratically. In this manner, the effective theory takes the form
\begin{eqnarray}\label{energy-10}
{\cal L}_{eff} &=& -\, \frac{1}{4} \, F_{\mu \nu }^2 - \frac{1}{2} \, G_\mu \, \frac{1}{{\left( {\Box  + m_Z^2} \right)}} \, G^\mu
\nonumber \\
&&
-\frac{1}{2} \, {G^\mu } \frac{{{\partial _\mu }{\partial _\nu }}}{{m_Z^2\left( {\Box  + m_Z^2} \right)}} \, {G^\nu } \; .
\end{eqnarray}
%
%
Whereas the field $G_\beta$ is given by
\begin{eqnarray}\label{energy-10a}
{G_\beta } &=&  - \frac{{\sin \theta _W {{\cos }^3} {{\theta _W}}}}{{4{\beta ^2}}}
\nonumber \\
&\times&{\partial ^\alpha }\left[ {\left( {{F_{\mu \nu }}{F^{\mu \nu }}} \right){F_{\alpha \beta }} + \left( {{F_{\mu \nu }}{{\tilde F}^{\mu \nu }}} \right){{\tilde F}_{\alpha \beta }}} \right] \; .
\end{eqnarray}
Furthermore, it can easily be seen that the effective model described by the Lagrangian (\ref{energy-10}) is a theory with non-local time derivatives. At this point, we again call attention to the fact that this section is aimed at studying the static potential, so that $\Box$ can be replaced by $-{\nabla ^2}$. For notational convenience we have maintained $\Box$, but it should be borne in mind that this paper essentially deals with the static case. Thus, the canonical quantization of this theory from the Hamiltonian point of view follows straightforwardly, as we will show below.
Now, if we wish to study quantum properties of the electromagnetic field in the presence of external electric and magnetic fields, we should split the ${F_{\mu\nu}}$-field strength as the sum of a classical background, $F_{0\mu\nu}$, and a small quantum fluctuation, $f_{\mu\nu}$, namely: $F_{\mu\nu}=f_{\mu\nu} + F_{0\mu\nu}$, as we have made previously. Therefore the Lagrangian density (\ref{energy-10}), up to quadratic terms in the fluctuations, is also expressed as
\begin{eqnarray}\label{energy-15}
{\cal L}_{eff}^{\,\,(2)} &=& - \, \frac{1}{4} \, {f_{\mu \nu }}\left[ {\frac{{\varpi \, \Box  + m_Z^2}}{{\Box  + m_Z^2}}} \right]{f^{\mu \nu }}
\nonumber \\
&&
\hspace{-1cm}
-\,2\,\delta F_{0\mu\nu}^2 \, {\tilde F}_{0}^{\;\,\rho\sigma} \, {\tilde F}_{0}^{\;\,\beta \lambda} \, {\partial^\alpha }{f_{\alpha \lambda }} \, \frac{1}{{\left( {\Box  + m_Z^2} \right)}} \, {\partial_\beta }{f_{\rho\sigma}}
\nonumber \\
&&
\hspace{-1cm}
-\,2\,\delta {\tilde F}_{0}^{\;\,\mu\nu} {\tilde F}_{0\alpha \lambda} {\tilde F}_{0}^{\;\,\rho\sigma} {\tilde F}_{0}^{\;\,\beta \lambda}
{\partial^\alpha}{f_{\mu\nu}} \, \frac{1}{{\left( {\Box  + m_Z^2} \right)}} \, {\partial _\beta }{f_{\rho\sigma}} \; , \hspace{0.7cm}
\end{eqnarray}
%
In this last expression, for notational simplicity, we have defined
\begin{eqnarray}
\varpi  = 1 - \delta \, (F_{0\mu\nu}F_{0}^{\;\,\mu\nu})^{2}
\;\; \mbox{and} \;\;
\delta  = \frac{\sin^2\theta _W \cos^6\theta _W}{16\beta^4} \; .
\;\;\;\;\;
\end{eqnarray}

It is to be specially noted that in the above Lagrangian density we have considered only the case of external magnetic fields, therefore we have ignored terms proportional to the external electric field.

In this way one encounters that the present effective theory provide us with a suitable starting point to study the interaction energy in the presence of an external magnetic field. To this end, we first consider the Hamiltonian framework of this new effective theory. The canonical momenta reads
\begin{eqnarray}\label{energy-20}
{\Pi ^\mu } &=&  - \, \left({\frac{{\varpi {\nabla ^2} - m_Z^2}}{{{\nabla ^2} - m_Z^2}}}\right) f^{0\mu}
\nonumber \\
&-& \!\!4\delta F_{0\rho\sigma}^2 {\tilde F}_{0}^{\;\,0\mu} {\tilde F}_{0}^{\;\,\beta \lambda} \, \frac{{{\partial_\beta }{\partial ^\alpha }}}{{\left( {{\nabla ^2} - m_Z^2} \right)}} \, {f_{\alpha \lambda }}
\nonumber \\
&-& \!\!4\delta {\tilde F}_{0}^{\;\,0\mu} {\tilde F}_{0\alpha \lambda} {\tilde F}_{0}^{\;\,\rho\sigma} {\tilde F}_{0}^{\;\,\beta \lambda}
\frac{{{\partial^\alpha }{\partial _\beta }}}{{\left( {{\nabla^2} - m_Z^2} \right)}} \, {f_{\rho\sigma}}
\nonumber \\
&-& \!\!4\delta {\tilde F}_{0}^{\;\,\rho\sigma} {\tilde F}_{0\alpha \lambda} {\tilde F}_{0}^{\;\,0\mu} {\tilde F}_{0}^{\;\,\beta \lambda}
\frac{{{\partial ^\alpha }{\partial _\beta }}}{{\left( {{\nabla ^2} - m_Z^2} \right)}} \, {f_{\rho\sigma}} \; .
\end{eqnarray}
This yields the usual primary constraint $\Pi^0 = 0$, and, while the canonical momentum is given by
\begin{eqnarray}
{\Pi ^i} = \left(\frac{{ {\varpi {\nabla ^2} - m_Z^2} }}{{ {{\nabla ^2} - m_Z^2}}} \right) {e^i} + 8\,\delta\, {\bf B}^{2}B^i B^j\frac{{{\partial _j}{\partial _k}}}{{ {{\nabla ^2} - m_Z^2}}} \, {e^k} \, . \; \;\;\;\;
\end{eqnarray}
Thus, the corresponding electric field due to the fluctuation takes the form
\begin{equation}
e_i = \frac{1}{{\det D}}\left[ {{\delta _{ij}}\det D + \frac{{\left( {{\bf B} \cdot \nabla } \right)}}{{{\Omega ^2}}}{ B_i}{ B_j}} \right]\frac{{\left( {{\nabla ^2} - m_Z^2} \right)}}{{\left( {\varpi {\nabla ^2} - m_Z^2} \right)}}{\Pi _j} \; ,
\end{equation}
where
\begin{eqnarray}
\det D = 1 + \frac{{{{\left( {{\bf B} \cdot \nabla } \right)}^2}}}{{{\Omega ^2}}}
\hspace{0.3cm} \mbox{and} \hspace{0.3cm}
\frac{1}{{{\Omega ^2}}} = \frac{{8\delta {{\bf B}^2}}}{{ {\varpi {\nabla ^2} - m_Z^2}}} \, . \;\;\;
\end{eqnarray}
Here, ${\bf B}$ represents the external (background) magnetic field around which the $a^{\mu}$-field fluctuates.
The canonical Hamiltonian is then
\begin{eqnarray}
{H_C} &=& \int {{d^3}x} \left[ {{\Pi _i}{\partial ^i}{a_0} - \frac{1}{2} \, {\Pi ^i}\frac{{\left( {{\nabla ^2} - m_Z^2} \right)}}{{\left( {\varpi {\nabla ^2} - m_Z^2} \right)}}{\Pi _i}} \right]
\nonumber \\
&+&\frac{{{{\bf B}^2}}}{2}\int {{d^3}x} \frac{{{{\left( {{\bf B} \cdot \nabla } \right)}^2}}}{{{{\left( {\Omega \det D} \right)}^2}}}{\partial _j}{\Pi _j}\frac{{\left( {{\nabla ^2} - m_Z^2} \right)}}{{\left( {\varpi {\nabla ^2} - m_Z^2} \right)}}{\partial _k}{\Pi _k} \nonumber\\
&-&4\,\delta\, {{\bf B}^2}  \int {{d^3}x}\frac{{{\partial _j}{\Pi _j}}}{{\det D}}\frac{{\left( {{\nabla ^2} - m_Z^2} \right)}}{{{{\left( {\varpi {\nabla ^2} - m_Z^2} \right)}^2}}}\left( {{\bf B} \cdot \nabla } \right)\nonumber\\
&\times&\left( {{\bf B} \cdot {\bf \Pi} } \right) \nonumber\\
&-&4\,\delta\, {{\bf B}^2}\int {{d^3}x} \frac{{{\partial _j}{\Pi _j}}}{{\det D}}\frac{{\left( {{\nabla ^2} - m_Z^2} \right)}}{{{{\left( {\varpi {\nabla ^2} - m_Z^2} \right)}^2}}}\left( {{\bf B} \cdot \nabla } \right)\nonumber\\
&\times&\left( {{\bf B} \cdot {\bf \Pi} } \right)
\nonumber\\
&+&4\,\delta\, {{\bf B}^4}\int {{d^3}x} \frac{{{\partial _j}{\Pi _j}}}{{\det D\left( {\varpi {\nabla ^2} - m_Z^2} \right)}}\frac{{{{\left( {{\bf B} \cdot \nabla } \right)}^2}}}{{{\Omega ^2}\det D}}\nonumber\\
&\times&\frac{{\left( {{\nabla ^2} - m_Z^2} \right)}}{{\left( {\varpi {\nabla ^2} - m_Z^2} \right)}}{\partial _k}{\Pi _k}\nonumber\\
&-& 8\,\delta \int {{d^3}x} \left( {{\bf B} \cdot \nabla } \right)\left( {{\bf B} \cdot {\bf \Pi} } \right)\frac{{\left( {{\nabla ^2} - m_Z^2} \right)}}{{{{\left( {\varpi {\nabla ^2} - m_Z^2} \right)}^2}}} \nonumber\\
&\times&\left( {{\bf B} \cdot \nabla } \right)\left( {{\bf B} \cdot {\bf \Pi} } \right)\nonumber\\
&-&8\,\delta\, {{\bf B}^2}\int {{d^3}x} \frac{{\left( {{\bf B} \cdot \nabla } \right)}}{{{\Omega ^2}}}\left( {{\bf B} \cdot {\bf \Pi} } \right)\frac{{\left( {{\nabla ^2} - m_Z^2} \right)}}{{{{\left( {\varpi {\nabla ^2} - m_Z^2} \right)}^2}}}\nonumber\\
&\times&{\left( {{\bf B} \cdot \nabla } \right)^2}{\partial _j}{\Pi _j}\nonumber\\
&-& 8\,\delta\, {{\bf B}^2}\int {{d^3}x} \frac{{{{\left( {{\bf B} \cdot \nabla } \right)}^2}}}{{{\Omega ^2}}}{\partial _j}{\Pi _j}\frac{{\left( {{\nabla ^2} - m_Z^2} \right)}}{{{{\left( {\varpi {\nabla ^2} - m_Z^2} \right)}^2}}}\nonumber\\
&\times&\left( {{\bf B} \cdot \nabla } \right)\left( {{\bf B} \cdot {\bf \Pi} } \right) \nonumber\\
&-& 8\,\delta\, {{\bf B}^4}\int {{d^3}x} \frac{{{{\left( {{\bf B} \cdot \nabla } \right)}^2}}}{{{\Omega ^2}}}{\partial _j}{\Pi _j}\frac{{{{\left( {{\bf B} \cdot \nabla } \right)}^2}}}{{{\Omega ^2}}}{\partial _k}{\Pi _k} \; .
\label{energy-25}
 \end{eqnarray}

The consistency condition $\dot{\Pi}_0=0$ leads to the secondary constraint ${\Gamma _1} \equiv {\partial _i}{\Pi ^i} = 0$ (Gauss's law), and together displays the first-class structure of the theory. The extended Hamiltonian that generates translations in time then reads
\begin{eqnarray}
H = H_C + \int {d^3 } x\left[ {c_0 \left( x \right) \Pi _0 \left( x \right) + c_1
\left( x\right)\Gamma _1 \left( x \right)} \right] \; ,
\end{eqnarray}
where $c_0 \left( x\right)$ and $c_1 \left( x \right)$ are the Lagrange multipliers.
Since $\Pi^0 = 0$ for all time and
\begin{eqnarray}
\dot{a}_0 \left( x \right)= \left[ \, {a_0\left( x \right) \, , \, H} \, \right] = c_0 \left( x \right) \; ,
\end{eqnarray}
which is completely arbitrary, we discard $a^0 $ and $\Pi^0$ because they add nothing to the description of the theory. Then, the extended Hamiltonian takes the form
\begin{eqnarray}\label{energy-30}
H &=& \int {{d^3}x} \left\{ {c\left( x \right){\partial _i}{\Pi ^i} - \frac{1}{2} \, {\Pi ^i}\frac{{\left( {{\nabla ^2} - m_Z^2} \right)}}{{\left( {\varpi {\nabla ^2} - m_Z^2} \right)}} \, {\Pi ^i}} \right\}
\nonumber \\
&+& \mbox{plus the other terms of the Eq.} \, (\ref{energy-25}) \, ,
\end{eqnarray}
where we have defined $c(x) = c_1 (x) - a_0 (x)$.
In accordance with the Dirac method, we must fix the gauge, that together with the first class constraint, ${\Gamma _1}(x)$, the full set of constraints become second class. A particularly convenient gauge-fixing condition is \cite{Gaete97}:
 \begin{equation}\label{energy-35}
\Gamma _2 \left( x \right) \equiv \int\limits_{C_{\zeta x} } {dz^\nu }
\, a_\nu\left( z \right) \equiv \int\limits_0^1 {d\lambda \, x^i } a_i \left( {
\lambda x } \right) = 0 \; ,
\end{equation}
where $\lambda$ $(0\leq \lambda\leq1)$ is the parameter describing the
space-like straight path $x^i = \zeta ^i + \lambda \left( {x - \zeta}
\right)^i $ , and $\zeta^{i}$ is a fixed point (reference point).
There is no essential loss of generality if we restrict our considerations to $\zeta^i=0$.
By means of this procedure, we arrive at the only non-vanishing equal-time Dirac bracket for
the canonical variables
\begin{eqnarray}\label{energy-40}
\left\{ {a_i \left( {\bf x} \right),\Pi ^j \left( {\bf y} \right)} \right\}^{\ast} &=& \delta_{i}^{\;\,j} \, \delta ^{\left( 3 \right)} \left( {{\bf x} - {\bf y}} \right)
\nonumber \\
\!&-&\! \partial_i^x
\int\limits_0^1 {d\lambda \, x^j } \delta ^{\left( 3 \right)} \left( {\lambda
{\bf x}- {\bf y}} \right) \; .
\end{eqnarray}

We may now proceed to determine the interaction energy for the effective theory under consideration. Recalling again that to do this we will work out the expectation value of the energy operator $H$ in the physical state $\left| \Phi  \right\rangle$. In that case we consider the stringy gauge-invariant state
\begin{eqnarray}\label{energy-45}
\left| \Phi  \right\rangle  &\equiv& \left|\, {\overline{\Psi} \left( {\bf y} \right)\Psi \left( {{{\bf y}^ {\prime} }} \right)} \,\right\rangle  \nonumber\\
&=& \overline{\Psi} \left( {\bf y} \right)\exp \left( {iq\int_{{{\bf y}^ {\prime} }}^{\bf y} {d{z^i}{a_i}\left( z \right)} } \right)\Psi \left( {{{\bf y}^ {\prime} }} \right)\left| 0 \right\rangle \; ,
\end{eqnarray}
where the line integral is along a space-like path on a fixed time slice, $q$ is the fermion charge and $\left| 0 \right\rangle$ is the physical vacuum state.
Next, taking into account the preceding Hamiltonian analysis, we then easily verify that
\begin{eqnarray}\label{energy-50}
{\Pi _i}\left( {\bf x} \right)\left| {\overline{\Psi} \left( {\bf y} \right)\Psi \left( {{{\bf y}^ {\prime} }} \right)} \right\rangle
&=& \overline{\Psi} \left( {\bf y} \right)\Psi \left( {{{\bf y}^ {\prime} }} \right){\Pi _i}\left( {\bf x} \right)\left| 0 \right\rangle
\nonumber \\
 &+& q\int_{\bf y}^{{{\bf y}^ {\prime} }} \! {d{z_i} \, {\delta ^{\left( 3 \right)}}\left( {{\bf z} - {\bf x}} \right)\left| \Phi  \right\rangle } \; .
\hspace{0.5cm}
\end{eqnarray}
Thus, the expectation value, ${\left\langle H \right\rangle _\Phi }$, simplifies to
\begin{equation}\label{energy-55}
{\left\langle H \right\rangle _\Phi } = {\left\langle H \right\rangle _0} + \left\langle H \right\rangle _\Phi ^{\left( 1 \right)} \; ,
\end{equation}
where ${\left\langle H \right\rangle _0} = \left\langle 0 \right|H\left| 0 \right\rangle$, whereas the $\left\langle H \right\rangle _0^{\left( 1 \right)}$ term is given by
\begin{equation}\label{energy-60}
\left\langle H \right\rangle _\Phi ^{\left( 1 \right)} =  - \frac{1}{2}\left\langle \Phi  \right|\int {{d^3}x} \, {\Pi ^i} \, \frac{{\left( {{\nabla ^2} - m_Z^2} \right)}}{{\left( {\varpi {\nabla ^2} - m_Z^2} \right)}} \, {\Pi _i}\left| \Phi  \right\rangle \; ,
\end{equation}

Following our earlier procedure \cite{Gaete97}, the static potential profile for two opposite charges located at ${\bf y}$ and ${\bf y}^{\prime}$ then reads
\begin{equation} \label{energy-65}
V(L) =  - \frac{{{q^2}}}{{4\pi \varpi}}\frac{{{e^{ - ML}}}}{L} + \frac{{{q^2}m_Z^2}}{{8\pi \varpi }} \, L \, \ln \left( {1 + \frac{{{\Lambda ^2}}}{{{M^2}}}} \right) \; ,
\end{equation}
where
\begin{eqnarray}
M = \frac{m_Z}{ \sqrt{1- \frac{\sin\theta_{W}\cos^3\theta_{W}}{8}\,\frac{{\bf B}^4}{\beta^4} } } \; ,
\end{eqnarray}
$|{\bf y} - {{\bf y}^ {\prime} }| \equiv L$, and $\Lambda$ is a cutoff. The $M$-parameter constraints the condition
\begin{eqnarray}
\frac{|{\bf B}|}{\beta} < 2.22
\; ,
\end{eqnarray}
which is consistent with our approximation in eq. (\ref{LBI}). Interestingly, the above static potential profile is analogous to that encountered for the electroweak sector of the Standard Model with an anomalous triple gauge boson couplings \cite{Gaete_AHEP_2021}. Finally, we would like to recall how to give a meaning to the cutoff $\Lambda$. To do that, we should remind \cite{Gaete_AHEP_2021} that our effective model for the electromagnetic field is an effective description that comes out upon integration over the $Z_{\mu}$-field, whose excitation is massive. $\ell_{Z}=m_{Z}^{-1}$, the Compton wavelength of this excitation, naturally defines a correlation distance. Physics at distances of the order or lower than $m_{Z}^{-1}$ must necessarily take into account a microscopic description of the $Z$-fields. This means that, if we work with energies of the order or higher than $m_{Z}$, our effective description with the integrated effects of $Z$ is no longer sensible. So, it is legitime that, for the sake of our analysis, we identify $\Lambda$ with $m_{Z}$. Thus, finally we end up with the following static potential profile:
\begin{equation}\label{energy-70}
V(L) =  - \frac{{{q^2}}}{{4\pi \varpi}}\frac{{{e^{ - ML}}}}{L} + \frac{{{q^2}m_Z^2}}{{8\pi \varpi }} \, L \, \ln \left( {1 + \frac{{{m_{Z} ^2}}}{{{M^2}}}} \right) \; .
\end{equation}
As already expressed in the Introduction, similar forms of interaction potentials have been reported before in the context of a theory of antisymmetric tensor fields that results from the condensation of topological defects as a consequence of the Julia-Toulouse mechanism \cite{Gaete2004},  gluodynamics in curved space-time \cite{Gaete2008}, a Higgs-like model \cite{Gaete2009},  Podolsky-axionic electrodynamics \cite{Gaete2012}, and a minimal supersymmetric extension of the Carroll-Field-Jackiw model for electrodynamics with a Chern-Simons-like Lorentz-symmetry violating term \cite{Gaete2013}.
%

%

%
The force associated with the previous potential is
\begin{equation}\label{force}
F(L) = -\frac{{{q^2}}}{{4\pi \varpi}}\left(\frac{1+ML}{L^2}\right)e^{ - ML} - \frac{{{q^2}m_Z^2}}{{8\pi \varpi }} \, \ln \left( {1 + \frac{{{m_{Z} ^2}}}{{{M^2}}}} \right) \; .
\end{equation}
Using the condition $\beta \gg |{\bf B}|$, the potential and the force are, approximately, given by
\begin{eqnarray}\label{Vapprox}
V(L) &\simeq&  - \frac{{{q^2}}}{{4\pi}} \frac{{{e^{ - m_{Z}L}}}}{L} + \frac{{{q^2}m_Z^2}}{{8\pi}} \, L \, \ln2
+{\cal O}\left(\frac{{\bf B}^4}{\beta^4}\right)
\; , \;
\nonumber \\
F(L) &\simeq&  -\frac{q^2}{4\pi L^2}\left( 1+m_{Z}L \right) \, e^{-m_{Z}L}
\nonumber \\
&&
-\frac{q^2}{8\pi} \, m_{Z}^2\, \ln2
+ {\cal O}\left(\frac{{\bf B}^4}{\beta^4}\right) \; .
\end{eqnarray}
Using the fundamental charge $\alpha=e^2=137^{-1}$ in natural units, $m_{Z}=91 \, \mbox{GeV}$, and a distance scale of
$L=10^{4} \, \mbox{GeV}^{-1}$, we obtain the force
\begin{eqnarray}\label{Fz}
F \, \simeq \, - \, 1.353636  \times  10^{10} \, \, \mbox{N} \; .
\end{eqnarray}
For the case of the attractive force of $W^{+} \, W^{-}$, we take $\lambda_{W} \sim \lambda_{Z} \sim 10^{-18} \, \mbox{m}$,
so we choose the distance scale at $L=0.1 \, \mbox{GeV}^{-1}$, and the correspondent force is (\ref{Fz}) up to fourth decimal place.
%
%


%
%

The expression above for the attractive interparticle force is valid
for distances larger than the Compton wavelength of the $Z$-gauge
boson $(\lambda_{Z} \sim 10^{-2} \, \mbox{GeV}^{-1} \sim 10^{-3} \, \mbox{fm})$, as previously pointed out.
So, to get an estimate of the attractive force between an electron-
positron pair, we have to consider distances that are also larger than
the electron's Compton wavelength. If we consider the interparticle
distance at order $10^{3} \, \mbox{fm}$, the force comes out of the order of (\ref{Fz})  .
On the other hand, if we wish to estimate the attraction between
a $W^{+} \, W^{-}$ pair, it is allowed to take $L \sim 10^{-2} \, \mbox{fm}$.
At such a distance, the force is estimated to be the result (\ref{Fz}) approximately.

%
%



\section{Conclusions and Final Remarks}
\label{sec6}
The introduction of a non-linear Born-Infeld (BI) theory in the Abelian sector of the electroweak model is proposed in this paper.
After the spontaneous symmetry breaking mechanism, the non-linearity yields anomalous couplings between the $Z$-boson and the photon in the approximation at the first order into the BI parameter. These couplings produce new effects beyond the Standard Model, like the decay $Z \rightarrow 3\,\gamma$.
Using the ATLAS upper bound $\mbox{Br}(Z \rightarrow 3\,\gamma) < 2.2 \times 10^{-6}$, we obtain the lower bound
$\sqrt{\beta} > 78.62 \, \mbox{GeV}$. In addition, through the recent results from the CMS and TOTEM Collaborations related to quartic photon self-couplings  \cite{CMSTOTEM}, we obtain stronger limits on the BI parameter, namely, $ \sqrt{\beta} > 502.84 \, \mbox{GeV}$ and
$\sqrt{\beta} > 591.42 \, \mbox{GeV}$. These latter bounds encompass our result $\sqrt{\beta} > 78.62 \, \mbox{GeV}$ and the previous bounds obtained in refs. \cite{PdeFabrittis} and  \cite{Ellis_PRL_2017}.
%

%
%

%
Subsequently, we investigate the kinetic aspects of this model when it is submitted to an external (and uniform) magnetic field ${\bf B}$.
We obtain the dispersion relations (DRs) of $Z$-boson and photon under this magnetic field, in the approximation $\beta \gg |{\bf B}|$.
The result (\ref{omega3ap}) and (\ref{omega4ap}) show the DRs depending on the direction of the magnetic field with the wave vector ${\bf k}$,
in which the second one fixes the condition on the BI parameter :
$\beta > 0.29 \sqrt{(({\bf B}\times {\bf k})^{2}-m_{Z}^2{\bf B}^2)/({\bf k}^2+m_{Z}^2)} $, respectively, with the wave vector constrained by $|\hat{{\bf B}}\times {\bf k}| > 91$ GeV. This condition implies into $\lambda < 1.38 \times 10^{-17}\, \mbox{m}$ when the magnetic field is perpendicular to the
$Z$-wave propagation direction.
Finally, within the gauge-invariant but path-dependent variables formalism, we have considered the confinement versus screening issue for our effective model, that is, a gauge theory with an anomalous coupling $3\gamma - Z$ under an external magnetic field. Again, a correct identification of physical degrees freedom has been fundamental for understanding the physics hidden in gauge theories. As we have shown, the interaction energy contains a linear potential, leading to the confinement of static charges.



In this contribution, we have been bound to a Born-Infeld-like approach to the weak
hypercharge factor of the underlying electroweak symmetry. As a consequence, only anomalous
$4$-gauge couplings between the photon and the $Z$-boson come out in the spontaneously
broken phase driven by the Higgs; anomalous couplings involving the $W^{\pm}$ gauge
mediators are left aside. In a forthcoming paper, based on a non-Abelian formulation of the
Born-Infeld theory \cite{Brain,Sevrin,Lombardo} – in our case, an $SU(2)$-Born-Infeld action to the fourth-
order in the field strength – we shall investigate aspects of anomalous couplings involving both
the charged and neutral gauge bosons in connection with dimension-$8$ operators. We are
motivated to focus on this activity by the understanding that the theoretical and
phenomenological analysis of anomalous electroweak gauge self-couplings, and the
experimental search for their effects, is a viable path to seek for imprints of new physics Beyond
Standard Model. With the high-energy, high-luminosity linear $e^{+} \, e^{-}$ colliders expected to be in
operation in the next decades, like the International Linear Collider (ILC), Compact Linear
Collider (CLIC) and the future muon collider, a new generation of high-precision electroweak
tests will be exploited and the effects of the anomalous couplings may constitute a relevant topic
in the agenda of this whole family of lepton colliders.
%
%

%
%

%
%

%
\section*{Acknowledgments}

We are grateful to Cristian Baldenegro for kind correspondence and fruitful discussion on anomalous couplings.
M. J. Neves thanks CNPq (Conselho Nacional de Desenvolvimento Cient\' ifico e Tecnol\'ogico), Brazilian scientific support federal agency, for partial financial support, Grant number 313467/2018-8. L.P.R. Ospedal is supported by the Ministry for Science, Technology and Innovations (MCTI) and CNPq under the Institutional Qualification Program (PCI). P. Gaete was partially supported by Fondecyt (Chile) grant 1180178 and by ANID PIA / APOYO AFB180002.

%
%

\end{document}